\documentclass[preprint,aps,amsmath,superscriptaddress,nofootinbib,tightenlines,showpacs]{revtex4}
\usepackage{txfonts}
\usepackage{mathrsfs}
\usepackage{amsmath}
\usepackage{amsfonts}
\usepackage{amssymb}
\usepackage{latexsym}
\usepackage{graphicx}
\usepackage{bm}
\usepackage{epsfig}
\usepackage[english]{babel}
\def \be {\begin{equation}}
\def \ee {\end{equation}}
\def \bea {\begin{eqnarray}}
\def \eea {\end{eqnarray}}
\def \nn {\nonumber}

\def \a {\alpha}
\def \b {\beta}

\def \d {\delta}

\def \m {\mu}
\def \n {\nu}
\def \k {\kappa}

\def \s {\sigma}
\def \r {\rho}
\def \o {\omega}

\def \th {\theta}
\def \Th {\Theta}

\def \t {\tau}
\def \dag {\dagger}
\def \p {\partial}

\def\bd{\begin{document}}
\def\ed{\end{document}}
\def\nn{\nonumber}
\def\bea{\begin{eqnarray}}
\def\eea{\end{eqnarray}}
\let\bm=\bibitem
\let\la=\label

\def\N{{\cal N}}
\def\sst{\scriptscriptstyle}
\def\thetabar{\bar\theta}
\def\Tr{{\rm Tr}}
\def\one{\mbox{1 \kern-.59em {\rm l}}}

\def\a{\alpha}      \def\da{{\dot\alpha}}
\def\b{\beta}       \def\db{{\dot\beta}}
\def\c{\gamma}  \def\C{\Gamma}  \def\cdt{\dot\gamma}
\def\d{\delta}  \def\D{\Delta}  \def\ddt{\dot\delta}
\def\e{\epsilon}        \def\vare{\varepsilon}
\def\f{\phi}    \def\F{\Phi}    \def\vvf{\f}
\def\h{\eta}
\def\k{\kappa}
\def\l{\lambda} \def\L{\Lambda}
\def\m{\mu} \def\n{\nu}
\def\o{\omega}
\def\P{\Pi}
\def\r{\rho}
\def\s{\sigma}  \def\S{\Sigma}
\def\t{\tau}
\def\th{\theta} \def\Th{\Theta} \def\vth{\vartheta}
\def\X{\Xeta}
\def\z{\zeta}
\def\w{\wedge}
\def\u{\underline}
\def\hs{\hspace}

\def\cA{{\cal A}} \def\cB{{\cal B}} \def\cC{{\cal C}}
\def\cD{{\cal D}} \def\cE{{\cal E}} \def\cF{{\cal F}}
\def\cG{{\cal G}} \def\cH{{\cal H}} \def\cI{{\cal I}}
\def\cJ{{\cal J}} \def\cK{{\cal K}} \def\cL{{\cal L}}
\def\cM{{\cal M}} \def\cN{{\cal N}} \def\cO{{\cal O}}
\def\cP{{\cal P}} \def\cQ{{\cal Q}} \def\cR{{\cal R}}
\def\cS{{\cal S}} \def\cT{{\cal T}} \def\cU{{\cal U}}
\def\cV{{\cal V}} \def\cW{{\cal W}} \def\cX{{\cal X}}
\def\cY{{\cal Y}} \def\cZ{{\cal Z}}

\def\ua{\underline{\alpha}} \def\ubb{\underline{\beta}}
\def\ug{\underline{\gamma}}
\def\ub{\underline{\phantom{\alpha}}\!\!\!\beta}
\def\uc{\underline{\phantom{\alpha}}\!\!\!\gamma}
\def\um{\underline{\mu}} \def\un{\underline{\nu}}
\def\ud{\underline\delta}
\def\ue{\underline\epsilon}
\def\una{\underline a}\def\unA{\underline A}
\def\unb{\underline b}\def\unB{\underline B}
\def\unc{\underline c}\def\unC{\underline C}
\def\und{\underline d}\def\unD{\underline D}
\def\une{\underline e}\def\unE{\underline E}
\def\unf{\underline{\phantom{e}}\!\!\!\! f}\def\unF{\underline F}
\def\unm{\underline m}\def\unM{\underline M}
\def\unn{\underline n}\def\unN{\underline N}
\def\unp{\underline{\phantom{a}}\!\!\! p}\def\unP{\underline P}
\def\unq{\underline{\phantom{a}}\!\!\! q}
\def\unQ{\underline{\phantom{A}}\!\!\!\! Q}
\def\unH{\underline{H}}
\def\ul{\underline}

\def\As {{A \hspace{-6.4pt} \slash}\;}
\def\bs {{b \hspace{-6.4pt} \slash}\;}
\def\Ds {{D \hspace{-6.4pt} \slash}\;}
\def\ds {{\del \hspace{-6.4pt} \slash}\;}
\def\ss {{\s \hspace{-6.4pt} \slash}\;}
\def\ks {{ k \hspace{-6.4pt} \slash}\;}
\def\ps {{p \hspace{-6.4pt} \slash}\;}
\def\pas {{{p_1} \hspace{-6.4pt} \slash}\;}
\def\pbs {{{p_2} \hspace{-6.4pt} \slash}\;}

\def\Fh{\hat{F}}
\def\Vh{\hat{V}}
\def\Xh{\hat{X}}
\def\ah{\hat{a}}
\def\xh{\hat{x}}
\def\yh{\hat{y}}
\def\ph{\hat{p}}
\def\xih{\hat{\xi}}

\def\psit{\tilde{\psi}}
\def\Psit{\tilde{\Psi}}
\def\tht{\tilde{\th}}

\def\At{\tilde{A}}
\def\Qt{\tilde{Q}}
\def\Rt{\tilde{R}}
\def\Nt{\tilde{N}}

\def\at{\tilde{a}}
\def\st{\tilde{s}}
\def\ft{\tilde{f}}
\def\pt{\tilde{p}}
\def\qt{\tilde{q}}
\def\vt{\tilde{v}}
\def\nt{\tilde{n}}

\def\delb{\bar{\partial}}
\def\bz{\bar{z}}
\def\bD{\bar{D}}
\def\bB{\bar{B}}

\def\bk{{\bf k}}
\def\bl{{\bf l}}
\def\bp{{\bf p}}
\def\bq{{\bf q}}
\def\br{{\bf r}}
\def\bx{{\bf x}}
\def\by{{\bf y}}
\def\bR{{\bf R}}
\def\bV{{\bf V}}

\def\d{\delta}\def\D{\Delta}\def\ddt{\dot\delta}

\def\p{\partial} \def\del{\partial}
\def\xx{\times}
\def\uno{\mbox{1 \kern-.59em {\rm l}}}

\def\trp{^{\top}}
\def\inv{^{-1}}
\def\dag{{^{\dagger}}}
\def\pr{\prime}

\def\rar{\rightarrow}
\def\lar{\leftarrow}
\def\lrar{\leftrightarrow}

\def\cw{{\cal W}}
\def\cz{{\cal Z}}
\def\tcm{\tilde{\cal M}}
\def\sgn{{\rm sgn}}
\def\sd {d^{4|4}}
\def\lan{\langle}
\def\ran{\rangle}

\def\tr{\mbox{tr}}
\def\sign{\mbox{sign}}
\def\fnl{f_\text{NL}}
\def\horava{Ho\v{r}ava}
\def\la{\langle}
\def\ra{\rangle}
\def\mb{\mathbf}
\def\nn{\nonumber}
\def\hl{Ho\v{r}ava-Lifshitz}
\def\p{\partial}
\def\dij{\delta_{ij}}
\def\tr{\mbox{tr}}
\def\sign{\mbox{sign}}
\def\fnl{f_\text{NL}}
\def\horava{Ho\v{r}ava}
\def\la{\langle}
\def\ra{\rangle}
\def\mb{\mathbf}
\def\nn{\nonumber}
\def\hl{Ho\v{r}ava-Lifshitz}
\def\p{\partial}
\def\dij{\delta_{ij}}

\begin{document}
\title{Static Charged Black Hole Solutions in Ho\v{r}ava-Lifshitz Gravity}
\author{Jin-Zhang Tang\footnote{Electronic address:JinzhangTang@pku.edu.cn}}
\affiliation{Department of Physics, and State Key Laboratory of
Nuclear Physics and Technology, Peking University, Beijing 100871,
China}

\date{\today\\ \vspace{1cm}}
\begin{abstract}
In the present work, we search static charged black hole solutions
to Ho\v{r}ava-Lifshitz gravity with or without projectability
condition. We consider the most general form of action which
electromagnetic field couples with Ho\v{r}ava-Lifshitz gravity. With
the projectability condition, we find (A)dS-Reissner-Nordstrom black
hole solution in Painlev\'e-Gullstrand type coordinates in the IR
region and a de-Sitter space-time solution in the UV region. Without
the projectability condition, in the IR region, we find an especial
static charged black hole solution.

\pacs{98.80.Cq}
\end{abstract}

\maketitle

\section{introduction}\label{sec-intro}
The \hl~ gravity which was introduced in
\cite{Horava:2008ih,Horava:2009uw} is intended to be a
power-counting renormalizable gravity theory. The basic idea behind
\horava's theory is that time and space may have different dynamical
scaling in UV limit. This was inspired by the development in quantum
critical phenomena in condensed matter physics, with the typical
model being Lifshitz scalar field theory\cite{Lifshitz,Chen:2009ka}.
In this \horava-Lifshitz theory, time and space will take different
scaling behavior as
\begin{equation}\label{scaling}
\mb{x}\rightarrow b\mb{x},\;\;\;\; t\rightarrow b^zt,
\end{equation}
where $z$ is the dynamical critical exponent characterizing the
anisotropy between space and time. Due to the anisotropy, instead of
diffeomorphism, we have the so-called foliation-preserving
diffeomorphism. The transformation is now just
 \bea
 t&\rightarrow& \tilde{t}(t), \nn\\
 x^i &\rightarrow& \tilde{x^i}(x^j,t).
 \eea
As a result, there is one more dynamical degree of freedom in
\hl-like gravity than in the usual general relativity. Such a degree
of freedom could play important role in UV physics, especially in
early cosmology\cite{Cai:2009dx,Chen:2009jr}. At IR, due to the
emergence of new gauge symmetry, this degree of freedom is not
dynamical any more such that the kinetic part of the theory recovers
the one of the general relativity.

Since time direction plays a privileged role in the whole
construction, it is more convenient to work with ADM metric
\begin{equation}\label{ADMmetric}
    ds^2=-N^2dt^2+h_{ij}(dx^i+N^idt)(dx^j+N^jdt),
\end{equation}
in which $N$ and $N_i$ are called ``lapse" and ``shift" variables
respectively.

Taking \hl~ gravity as a new gravitational theory, it is an
important issue to study its black hole solutions. In papers
\cite{Lu2009,Nastase2009,Kehagias:2009is,AhmadGhodsi2009,Colgain:2009fe,park2009,RongCai-2009,Hatefi2009},
it was assumed that the metric of the black solutions had the
following Schwarzschild coordinates form
\begin{equation}\label{Nr}
ds^2=-N(r)^{2}dt_{S}^2+\frac{dr^2}{g(r)}+r^{2}(d\theta^{2}+\sin^{2}\theta
d\phi^2).
\end{equation}
From this metric ansatz, it was found that there were new black hole
solutions, even at IR. For example, in \cite{Kehagias:2009is}, based
on a modified \hl~-type action, an asymptotically flat solution with
\be
 g=N^2=1+\omega r^2-\sqrt{r(\o^2r^3+4\o M)}
\ee was found. And paper \cite{RongCai-2009} find a new static
charged black hole solution in the IR region, paper
\cite{Hatefi2009} had studied the extremal rotating no-charged black
hole solutions. However, in the above ansatz (\ref{Nr}) the ``lapse
function'' $N(r)$ obviously breaks the ``projectability condition''
which means that $N$ only is the function of $t$. This is introduced
in \cite{Horava:2008ih,Horava:2009uw}. For the metric of the form
\eqref{Nr}, we can work in the Painlev\'e-Gullstrand coordinates by
making a transformation
\begin{equation}\label{coortransform01}
dt_{S}=dt_{PG}-\frac{\sqrt{1-N^{2}}}{N^{2}}dr.
\end{equation}
 Then the ansatz \eqref{Nr} becomes
\begin{equation}
ds^{2}=-dt_{PG}^{2}+(dr+\sqrt{1-N^{2}}dt_{PG})^{2}+(\frac{1}{g}-\frac{1}{N^{2}})dr^{2}+r^{2}(d\theta^{2}+\sin^{2}\theta
d\phi^2).
\end{equation}
Comparing with the ADM metric, we find that $N(t_{PG})=1$ which is
in accord with the projectability condition and if $g=N^{2}$ we
reach \eqref{ADMmetric}. For instance, a Reissner-Nordstrom black
hole after the transform \eqref{coortransform01} has the form
\begin{equation}
ds^{2}=-dt_{PG}^{2}+(dr\pm\sqrt{\frac{2GM}{r}-\frac{Q^{2}}{r^{2}}}dt_{PG})^{2}+r^{2}\left(d\theta^{2}+\sin^{2}\theta
d\phi^{2}\right).
\end{equation}
So paper \cite{static_black_hole_of_HL} search non-charge black hole
solutions to modified \hl~ gravity which was introduced by
\cite{Visser_2009} with the metric ansatz as
\begin{equation}\label{ansatz_N(t)_Nr}
ds^2=-N^{2}dt^2+\frac{1}{f(r)}(dr+N^{r}dt)(dr+N^{r}dt)+r^{2}(d\theta^{2}+\sin^{2}\theta
d\phi^2),
\end{equation}
where $N$ only is a function of $t$. Paper
\cite{static_black_hole_of_HL} found maximally symmetric space
solution with curvature $\Lambda_{W}$ and in the IR region, and
(A)dS-Schwarzschild black hole solution in Painlev\'e-Gullstrand
type coordinates was found. In the UV region, it found a de-Sitter
space-time solution.

Then papers \cite{Kiritsis_blackhole,Capasso_blackhole} search black
hole solutions with the same metric ansatz as
\eqref{ansatz_N(t)_Nr}, but $N$ only is the function of $r$. They
seem get some new black hole solutions. Actually the solutions of
paper \cite{Capasso_blackhole} in the IR region is very similar as
paper \cite{park2009}. Furthermore, paper \cite{Setare2009} had
studied the plane symmetric solutions in Ho\v{r}ava-Lifshitz theory.

In the present work, we search static charged black hole solutions
to \hl~ gravity coupling with electromagnetic field. We work on the
same metric ansatz as \eqref{ansatz_N(t)_Nr}. Whether the metric
ansatz respects the projectability condition is determined on
whether $N$ only is the function of $t$. With the projectability
condition, we find (A)dS-Reissner-Nordstrom black hole solution in
Painlev\'e-Gullstrand type coordinates in the IR region and a
de-Sitter space-time solution in the UV region. Without the
projectability condition, in the IR region, we find a static charged
black hole solution which is similar as the results of
\cite{RongCai-2009}. In the UV region, we find the same solution as
\cite{Lu2009}.

\section{The \hl~ gravity}\label{sec-modification}
In this section, we give a brief review of \hl~gravity. Using the
ADM formalism, the action of this \horava-Lifshitz gravitational
theory is given by\cite{Horava:2008ih,Horava:2009uw}

\begin{eqnarray}\label{S_g:origin}
\nn
S_{HL}&=&\int dtd^{3}\mb{x}(\mathcal{L}_{K}+\mathcal{L}_{V}),\\
\nn
\mathcal{L}_{K}&=&\sqrt{h}N\left\{\frac{2}{\kappa^2}(K_{ij}K^{ij}-\lambda
K^2)\right\},\\
\nn
\mathcal{L}_{V}&=&\sqrt{h}N\left\{\frac{\kappa^2\mu^2(\Lambda_{W}R-3\Lambda^2_{W})}{8(1-3\lambda)}+\frac{\kappa^2\mu^2(1-4\lambda)}{32(1-3\lambda)}R^2\right.\\
&&\left.-\frac{\kappa^2}{2\omega^4}\left(C_{ij}-\frac{\mu\omega^2}{2}R_{ij}\right)\left(C^{ij}-\frac{\mu\omega^2}{2}R^{ij}\right)\right\},
\end{eqnarray}
where $\mathcal{L}_{K}$ is the kinetic term and $\mathcal{L}_{V}$ is
the potential term. If we consider the term which represents a
``soft'' violation of the ``detailed balance'' condition in
\cite{Horava:2009uw} and add the term in the action, The kinetic
term $\mathcal{L}_{K}$ is the same, and the potential term
$\mathcal{L}_{V}$ becomes
\begin{eqnarray}
\nn
\mathcal{L}_{V}&=&\sqrt{h}N\left\{\frac{\kappa^2\mu^2(\Lambda_{W}R-3\Lambda^2_{W})}{8(1-3\lambda)}+\frac{\kappa^2\mu^2(1-4\lambda)}{32(1-3\lambda)}R^2\right.\\
&&\left.-\frac{\kappa^2}{2\omega^4}\left(C_{ij}-\frac{\mu\omega^2}{2}R_{ij}\right)\left(C^{ij}-\frac{\mu\omega^2}{2}R^{ij}\right)+\frac{\Omega\kappa^{2}\mu^{2}}{8(3\lambda-1)}R\right\}.\label{action_term_violation_debalance}
\end{eqnarray}
The last term has been introduced in
\cite{Horava:2009uw,Nastase2009,Kehagias:2009is}. In the action,
$\lambda,\kappa,\mu,\omega$,$\Lambda_{W}$ and $\Omega$ are the
coupling parameters, and $C_{ij}$ is the Cotton tensor defined by
\begin{equation}
C^{ij}=\epsilon^{ikl}\nabla_{k}\left(R^j_l-\frac{1}{4}R\delta^j_l\right).
\end{equation}

 The study of the perturbations around
the Minkowski vacuum shows that there is ghost excitation when
$\frac{1}{3}<\lambda<1$. This indicates that the theory is only
well-defined in the region $\l\leq \frac{1}{3}$ and $\l\geq 1$.
Since the theory should be RG flow to IR with $\l =1$, we expect
that at UV, $\l >1$ to have a well-defined RG flow. At IR, $\l =1$,
the kinetic term recovers the one of standard general relativity.
Comparing \eqref{S_g:origin} to the action of the general relativity
in the ADM formalism, the speed of light, the Newton's constant and
the cosmological constant emerge as
\begin{eqnarray}\label{c:former}
  c = \frac{\kappa^2\mu}{4}\sqrt{\frac{\Lambda_W}{1-3\lambda}},
  \hspace{3ex}
  G = \frac{\kappa^2}{32\pi c},
  \hspace{3ex}\Lambda=\frac{3}{2}\Lambda_W.
\end{eqnarray}
It follows from \eqref{c:former} that for $\lambda>1/3$ ,the
cosmological constant $\Lambda_{W}$ has to be negative. It was
noticed in \cite{Lu2009} that  if we make an analytic continuation
of the parameters
\begin{equation}
\mu \to i\mu, \hspace{4ex} \omega^2 \to -i\omega^2,
\end{equation}
the four-dimensional action \eqref{S_g:origin} remains real as
\begin{eqnarray}\label{S_g:continuation}
\nn
\mathcal{L}_{K}&=&\sqrt{h}N\left\{\frac{2}{\kappa^2}(K_{ij}K^{ij}-\lambda
K^2)\right\},\\
\nn
\mathcal{L}_{V}&=&\sqrt{h}N\left\{\frac{\kappa^2\mu^2(\Lambda_{W}R-3\Lambda^2_{W})}{8(3\lambda-1)}+\frac{\kappa^2\mu^2(1-4\lambda)}{32(3\lambda-1)}R^2\right.\\
&&\left.+\frac{\kappa^2}{2\omega^4}\left(C_{ij}-\frac{\mu\omega^2}{2}R_{ij}\right)\left(C^{ij}-\frac{\mu\omega^2}{2}R^{ij}\right)\right\}.
\end{eqnarray}
In this case, the emergent speed of light becomes
\begin{equation}
c= \frac{\kappa^2\mu}{4}\sqrt{\frac{\Lambda_W}{3\lambda-1}}.
\end{equation}
The requirement that this speed be real implies that $\Lambda_{W}$
must be positive for $\lambda>\frac{1}{3}$.

\section{Electromagnetic Field in \hl~ Static Curved Spacetime}
The electrodynamics in curved spacetime has been discussed many
times in the old days. Ellis(1973) first wrote down the Maxwell's
equations in $3+1$ congruence language. Later the Maxwell's
equations in $3+1$ form were fully discussed by Thorne, Price, and
Macdonald(1986). The Maxwell's equations in a curved spacetime could
be written as
\begin{eqnarray}
\label{eq-F01}
&&\nabla_{\nu}F^{\mu\nu}=0,\\
\label{eq-F02}
&&\partial_{\mu}F_{\nu\rho}+\partial_{\nu}F_{\rho\mu}+\partial_{\rho}F_{\mu\nu}=0,
\end{eqnarray}
where $F_{\mu\nu}$ is the antisymmetric electromagnetic tensor and
$F^{\mu\nu}=F_{\alpha\beta}g^{\alpha\mu}g^{\beta\nu}$. The electric
field $E_{i}$ and magnetic field $H_{i}$ are related to $F_{\mu\nu}$
as
\begin{equation}\label{EMfromF}
E_{i}=F_{ti},\;H_{i}=-\frac{\epsilon_{ijk}}{2}\sqrt{-g}F^{jk},
\end{equation}
where $E_{i}$ and $H_{i}$ are spatial vectors. The electromagnetic
action in the curved spacetime could be written in $3+1$ form
\begin{equation}
S_{em}=\int dtd^{3}\mb{x}\sqrt{h}Ng_{em}
\left[F_{tk}F^{tk}+F_{kt}F^{kt}+F_{ij}F^{ij}\right],
\end{equation}
where $g_{em}$ is a constant. But the action at the Lifshitz point
may be very different. This has been discussed by \horava
\cite{Horava_Gugue_field}, Chen and Huang
\cite{Chen_Huang_Lishitz_Point}. In the paper
\cite{Chen_Huang_Lishitz_Point}, Chen and Huang showed the action of
Yang-Mills gauge field at the Lifshitz point in the flat spacetime
as
\begin{equation}
S_{YM}=\frac{1}{2}\int
dtd^{d}\mb{x}\left[\frac{1}{g^{2}_{E}}Tr(E_{i}E_{i})-\sum_{J\geqslant
2}\mathcal{O}_{J}\star F^{J} \right],
\end{equation}
where
\begin{equation}
\mathcal{O}_{J}=\frac{1}{g^{J}_{E}}\sum^{n_{J}}_{n=0}(-1)^{n}\frac{\lambda_{J,n}}{M^{2n+\frac{d+1}{2}J-d-1}}D^{2n}.
\end{equation}
Here $F$ and $D$ are the abbreviated denotation for $F_{ij}$ and
$D_{k}$ respectively, and $\lambda_{J,n}$ are the the coupling with
zero energy dimension. Similarly $D^{2n}\star F^{J}$ also contains
all possible independent combinations of $D_{k}$ and $F_{ij}$. The
action of Yang-Mills theory at Lifshitz point in curved spacetime
should be similar with action show above. To the static charged
black hole, $F_{tr}$ component of $F_{\mu\nu}$ is not zero and it
must be functions of $r$. If there are independent magnetic charges
in the world and the static black hole absorbs some magnetic
charges, the magnetic field component $H_{r}$ should not be zero.
From \eqref{EMfromF}, $F^{\theta\phi}$ should not be zero. So the
action of electromagnetic field to static charged black hole will be
reduced to
\begin{equation}
\label{ActionOfEM} S_{em}=\int dtd^{3}\mb{x}\mathcal{L}_{em}=\int
dtd^{3}\mb{x}\sqrt{h}N
2g_{em}\left[F_{tr}F^{tr}+F_{\theta\phi}F^{\theta\phi}\right].
\end{equation}
From the equations \eqref{eq-F01} and \eqref{eq-F02} we get four
independent equations
\begin{equation}\label{eq-F0r01}
\partial_{r}\left(\sqrt{h}NF^{tr}\right)=0.
\end{equation}
\begin{equation}\label{eq-FThetaPhi01}
\partial_{\theta}\left(\sqrt{h}NF^{\theta\phi}\right)=0,\;\;\;\partial_{\phi}\left(\sqrt{h}NF^{\theta\phi}\right)=0.
\end{equation}
\begin{equation}\label{eq-FThetaPhi02}
\partial_{r}F_{\theta\phi}+\partial_{\phi}F_{r\theta}+\partial_{\theta}F_{\phi
r}=0.
\end{equation}

\section{Static Charged Black Hole Solutions}
As the discussion in the first paragraph, we now seek the static
charged black hole solutions with the metric ansatz
\begin{equation}\label{ansatz-sin}
ds^2=-N(t,r)^{2}dt^2+\frac{1}{f(r)}(dr+N^{r}dt)(dr+N^{r}dt)+r^{2}(d\theta^{2}+\sin^{2}\theta
d\phi^2).
\end{equation}

With this metric ansatz, from \eqref{eq-F0r01}, $F^{tr}$ should
satisfy the equation
\begin{equation}
\partial_{r}\left(\frac{N}{\sqrt{f}}r^{2}F^{tr}\right)=0.
\end{equation}
The solution of this equation is
\begin{equation}
F^{tr}=\frac{Q_{e}\sqrt{f}}{Nr^{2}},
\end{equation}
where $Q_{e}$ is an integration constant. From
\eqref{eq-FThetaPhi01} and \eqref{eq-FThetaPhi02}, $F^{\theta\phi}$
and $F_{\theta\phi}$ should satisfy
\begin{equation}
\partial_{\theta}\left(\frac{N}{\sqrt{f}}r^{2}\sin{\theta}F^{\theta\phi}\right)=0,\;\partial_{\phi}\left(\frac{N}{\sqrt{f}}r^{2}\sin{\theta}F^{\theta\phi}\right)=0,\;
\partial_{r}F_{\theta\phi}=0.
\end{equation}
The solution of the three equations are
\begin{equation}
F^{\theta\phi}=\frac{Q_{m}}{r^{4}\sin{\theta}},\;F_{\theta\phi}=Q_{m}\sin{\theta},
\end{equation}
where $Q_{m}$ is an integration constant. So from the action
\eqref{ActionOfEM}, the Lagrangian of electromagnetic field is
\begin{equation}\label{action-L_em}
\mathcal{L}_{em}=-2g_{em}\frac{N}{\sqrt{f}}\frac{Q_{e}^{2}-Q_{m}^{2}}{r^{2}}.
\end{equation}

The whole action of electromagnetic field couples with \hl~ gravity
is $S=S_{HL}+S_{em}$. Substituting the metric ansatz
\eqref{ansatz-sin} into the Lagrangians \eqref{S_g:origin} and
\eqref{ActionOfEM}, up to an overall scaling constant, we get
\begin{eqnarray}\label{L_KV}
\nn
\mathcal{L}_{K}=&&\frac{1}{N\sqrt{f}}\left\{(1-\lambda)r^{2}f^{2}\left(N^{'}_{r}+N_{r}\frac{f^{'}}{2f}\right)^{2}+2(1-2\lambda)f^{2}N_{r}^{2}\right.\\
&&\hspace{2ex}\left.-4\lambda rf^{2}N_{r}\left(N^{'}_{r}+N_{r}\frac{f^{'}}{2f}\right) \right\},\\
\nn
\mathcal{L}_{V}=&&\frac{N}{\sqrt{f}}\left\{2-3\Lambda_{W}r^2-2f-2rf^{'}+\frac{\lambda-1}{2\Lambda_{W}}f^{'2}\right.\\
&&\hspace{2ex}\left.-\frac{2\lambda(f-1)}{\Lambda_{W}r}f^{'}+\frac{(2\lambda-1)(f-1)^2}{\Lambda_{W}r^2}\right\},\\
\mathcal{L}_{em}=&&-2\tilde{g}_{em}\frac{N}{\sqrt{f}}\frac{Q_{e}^{2}-Q_{m}^{2}}{r^{2}}.
\end{eqnarray}
where $N_{r}=N^{r}/f$ and $'$ means derivative to $r$. Here we have
set $c=1$ and $\tilde{g}_{em}=\kappa^{2}g_{em}/2$. The full
Lagrangian is
$\mathcal{L}=\mathcal{L}_{K}+\mathcal{L}_{V}+\mathcal{L}_{em}$. By
varying the action with respect to the functions $N_{r}$ , $f$ and
$N(t)$, we obtain three equations of motions,
\begin{eqnarray}\label{By_Nr}
\nn
0=2(1-\lambda)r^{2}f^{2}\frac{1}{N}&&\left\{N_{r}^{''}+\frac{f^{''}}{2f}N_{r}+\frac{3}{2}\frac{f^{'}}{f}N_{r}^{'}+2\frac{N_{r}^{'}}{r}+\frac{1-2\lambda}{1-\lambda}\frac{f^{'}}{f}\frac{N_{r}}{r}-2\frac{N_{r}}{r^{2}}\right.\\
&&\left.-\frac{N^{'}}{N}\left(N_{r}^{'}+N_{r}\frac{f^{'}}{2f}-\frac{2\lambda}{1-\lambda}\frac{N_{r}}{r}\right)\right\},
\end{eqnarray}
\begin{eqnarray}\label{By_f}
\nn
0&&=-\left(\frac{f^{'}}{2f}\frac{1}{N}+\frac{N^{'}}{N^{2}}\right)\left\{(1-\lambda)r^{2}fN_{r}\left(N_{r}^{'}+N_{r}\frac{f^{'}}{2f}\right)-2\lambda
rfN_{r}^{2}\right\}\\
\nn
&&+\left(N^{'}-\frac{f^{'}}{2f}N\right)\left\{-2r+\frac{\lambda-1}{\Lambda_{W}}f^{'}-\frac{2\lambda(f-1)}{\Lambda_{W}r}\right\}+N\left\{\frac{\lambda-1}{\Lambda_{W}}f^{''}+\frac{2(1-\lambda)(f-1)}{\Lambda_{W}r^{2}}\right\}\\
\nn
&&+\frac{1}{N}\left\{(1-\lambda)r^{2}fN_{r}N_{r}^{''}+\frac{1}{2}(1-\lambda)r^{2}f^{''}N_{r}^{2}-(1-\lambda)r^{2}fN_{r}^{'2}+(1-\lambda)r^{2}f^{'}N_{r}N_{r}^{'}\right.\\
&&+\left.2(1+\lambda)rfN_{r}N_{r}^{'}+(1-\lambda)rf^{'}N_{r}^{2}+(6\lambda-4)fN_{r}^{2}\right\}+\frac{1}{2\sqrt{f}}\left\{\mathcal{L}_{K}+\mathcal{L}_{V}\right\}
-\tilde{g}_{em}\frac{N}{f}\frac{Q_{e}^{2}-Q_{m}^{2}}{r^{2}},
\end{eqnarray}
\begin{equation}\label{By_N}
0=-\mathcal{L}_{K}+\mathcal{L}_{V}+\mathcal{L}_{em}.
\end{equation}

\subsection{Solutions with Projectability Condition}
For the projectability condition, lapse function $N$ only is a
function of $t$, $N^{'}$ is zero and we can set $N(t)=1$ with the
time rescaling. Obviously for all $\lambda$, $N_{r}=0$ is a solution
of \eqref{By_Nr}. In this case, we assume ansatz $f(r)=1+yr^{2}$,
just when $Q=0$, from \eqref{By_f},\eqref{By_N}, we get quadratic
equations of $y$,
\begin{eqnarray}
&&y^{2}-2\Lambda_{W}y-3\Lambda_{W}^{2}=0,\\
&&y^{2}+2\Lambda_{W}y+\Lambda_{W}^{2}=0.
\end{eqnarray}
Their solution is $y=-\Lambda_{W}$. This solutions corresponds to a
maximally symmetric space with curvature $\Lambda_{W}$. This result
has been shown in paper \cite{static_black_hole_of_HL}.

In the IR region where $\lambda=1$, the equation \eqref{By_Nr} is
reduced to
\begin{equation}
\frac{f^{'}}{f}\frac{N_{r}}{r}=0.
\end{equation}
Its $N_{r}=0$ solution has been discussed above. It has another
solution as $f=constant$. when $f$ is a constant, we could get the
same equation from \eqref{By_f},\eqref{By_N} as
\begin{equation}
(N_{r}^{2})'+\frac{N_{r}^{2}}{r}+\frac{1}{2f^{2}}\left\{-3\Lambda_{W}r+\frac{2(1-f)}{r}+\frac{(1-f)^{2}}{\Lambda_{W}r^{3}}-2\tilde{g}_{em}\frac{Q_{e}^{2}-Q_{m}^{2}}{r^{3}}\right\}=0.
\end{equation}
The solutions of this equation are
\begin{equation}
N_{r}=\pm\frac{1}{f}\sqrt{\frac{M}{r}+\frac{\Lambda_{W}}{2}r^{2}+(f-1)+\frac{(1-f)^{2}}{2\Lambda_{W}r^{2}}-\tilde{g}_{em}\frac{Q_{e}^{2}-Q_{m}^{2}}{r^{2}}}.
\end{equation}
Especially when $f=1$, the solution reduces to
\begin{equation}\label{RN_blackhole}
N_{r}=\pm\sqrt{\frac{M}{r}+\frac{\Lambda_{W}}{2}r^{2}-\tilde{g}_{em}\frac{Q_{e}^{2}-Q_{m}^{2}}{r^{2}}}.
\end{equation}
It is a (A)dS-Reissner-Nordstrom black hole written in
Painlev\'e-Gullstrand type coordinates. Especially when
$Q_{e}=Q_{m}=0$, it is (A)dS-Schwarzschild black hole which has been
shown in \cite{static_black_hole_of_HL}. The electric field of the
black hole \eqref{RN_blackhole} is $E_{r}=F_{tr}=(-Q_{e})/r^{2}$.
Its charge is
\begin{equation}
Q_{BH}=\frac{1}{4\pi}\int_{S}\vec{E}\cdot
d\vec{\sigma}=\frac{1}{4\pi}E_{r}\cdot4\pi r^{2}=-Q_{e},
\end{equation}
where $S$ is a closed surface everywhere with the same $r$ and
$d\vec{\sigma}$ is surface integral element. We should choose
$\tilde{g}_{em}=1$ when the solution \eqref{RN_blackhole} is a
charged black hole with charge $\pm Q_{e}$.

In the UV region where $\lambda\neq1$, when $f$ is a constant, from
\eqref{By_Nr},\eqref{By_f} and \eqref{By_N}, we get three equations,
\begin{eqnarray}
0=&&N_{r}^{''}+2\frac{N_{r}^{'}}{r}-2\frac{N_{r}}{r^{2}},\label{By_Nr_lambda_neq_1}\\
\nn 0=&& (1-\lambda)r^{2}N_{r}^{'2}-4\lambda
rN_{r}N_{r}^{'}+2(1-2\lambda)N_{r}^{2}-\frac{4(1-\lambda)(f-1)}{\Lambda_{W}r^{2}f}\\
&&-\frac{1}{f^{2}}\left\{2(1-f)-3\Lambda_{W}r^{2}+\frac{(2\lambda-1)(f-1)^{2}}{\Lambda_{2}r^{2}}\right\}+2\tilde{g}_{em}\frac{Q_{e}^{2}-Q_{m}^{2}}{f^{2}r^{2}},\label{By_f_lambda_neq_1}\\
\nn 0=&& (1-\lambda)r^{2}N_{r}^{'2}-4\lambda
rN_{r}N_{r}^{'}+2(1-2\lambda)N_{r}^{2}\\
&&-\frac{1}{f^{2}}\left\{2(1-f)-3\Lambda_{W}r^{2}+\frac{(2\lambda-1)(f-1)^{2}}{\Lambda_{2}r^{2}}\right\}+2\tilde{g}_{em}\frac{Q_{e}^{2}-Q_{m}^{2}}{f^{2}r^{2}}.\label{By_N_lambda_neq_1}
\end{eqnarray}
Just when $f=1$ and $Q_{e}^{2}-Q_{m}^{2}=0$, they have a solution as
\begin{equation}
 N_{r}=\pm\;\sqrt{\frac{\Lambda_{W}}{(3\lambda-1)}}r.\label{solutions-lambda-neq1}
\end{equation}
This solution actually describes the same de-Sitter space-time. One
easy way to see this point is to change inversely into the
Schwarzschild coordinates. This result is the same as paper
\cite{static_black_hole_of_HL}.

\subsection{Solutions without Projectability Condition}
Without projectability condition, $N$ is function of $r$, $N'$ isn't
zero. In the IR region where $\lambda=1$, the equation \eqref{By_Nr}
is reduced to
\begin{equation}
\left(2\frac{N^{'}}{N}-\frac{f^{'}}{f}\right)\frac{N_{r}}{r}=0.
\end{equation}
Its solutions are $N_{r}=0$ or $N^{2}=f$. In the case $N_{r}=0$,
with the same discussion above, we find a solution
\begin{equation}
f=1-\Lambda_{W}r^{2},
\end{equation}
and the function $N(r)$ is unconstrained. This result is the same as
paper \cite{Lu2009}.

When $\lambda=1$, in the case $N^{2}=f$, from
\eqref{By_f},\eqref{By_N}, we get one same equation,
\begin{eqnarray}\label{eq_WithoutPrC_lambdaeq1_N2eqf}
0=2\left[rfN_{r}^{2}\right]^{'}+\left\{2(1-f)-3\Lambda_{W}r^{2}-2rf^{'}+\frac{2(1-f)}{\Lambda_{W}r}f^{'}+\frac{(f-1)^{2}}{\Lambda_{W}r^{2}}\right\}-2\tilde{g}_{em}\frac{Q_{e}^{2}-Q_{m}^{2}}{r^{2}}.
\end{eqnarray}
This equation has a solution as
\begin{equation}\label{chargedblackhole_without_PC_01}
N_{r}^{2}=\frac{\beta}{rf},\;
f=1-\Lambda_{W}r^{2}-\sqrt{cr+2\tilde{g}_{em}\Lambda_{W}\left(Q_{e}^{2}-Q_{m}^{2}\right)},
\end{equation}
Where $\beta,c$ are integration constants. If we reconsider the
action \eqref{action_term_violation_debalance} which contains the
``soft'' violation term, we could get a equation similar as
\eqref{eq_WithoutPrC_lambdaeq1_N2eqf} as
\begin{eqnarray}\label{eq_WithoutPrCDetBan_lambdaeq1_N2eqf}
0=2\left[rfN_{r}^{2}\right]^{'}+\left\{-2\frac{\Omega-\Lambda_{W}}{\Lambda_{W}}(1-f-rf^{'})-3\Lambda_{W}r^{2}+\frac{2(1-f)}{\Lambda_{W}r}f^{'}+\frac{(f-1)^{2}}{\Lambda_{W}r^{2}}\right\}-2\tilde{g}_{em}\frac{Q_{e}^{2}-Q_{m}^{2}}{r^{2}}.
\end{eqnarray}
Its solution is
\begin{equation}\label{chargedblackhole_without_PC_02}
N_{r}^{2}=\frac{\beta}{rf},\;
f=1+\left(\Omega-\Lambda_{W}\right)r^{2}-\sqrt{\Omega(\Omega-2\Lambda_{W})r^{4}+cr+2\tilde{g}_{em}\Lambda_{W}\left(Q_{e}^{2}-Q_{m}^{2}\right)},
\end{equation}
Where $\beta,c$ are integration constants. When $Q_{m}=0$, this
solution is similar as paper \cite{RongCai-2009} and When
$Q_{e}^{2}-Q_{m}^{2}=0$ it is the similar as \cite{park2009}. The
solutions
\eqref{chargedblackhole_without_PC_01},\eqref{chargedblackhole_without_PC_02}
are two especial static charged black hole solutions.

In the UV region where $\lambda\neq1$, when $N_{r}=0$, from
\eqref{By_f}.\eqref{By_N}, we get two equations,
\begin{eqnarray}
0=&&2-3\Lambda_{W}r^2-2f-2rf^{'}+\frac{\lambda-1}{2\Lambda_{W}}f^{'2}
-\frac{2\lambda(f-1)}{\Lambda_{W}r}f^{'}+\frac{(2\lambda-1)(f-1)^2}{\Lambda_{W}r^2}-2\tilde{g}_{em}\frac{Q_{e}^{2}-Q_{m}^{2}}{r^{2}},\\
0=&&\left(N^{'}-\frac{f^{'}}{2f}N\right)\left\{-2r+\frac{\lambda-1}{\Lambda_{W}}f^{'}-\frac{2\lambda(f-1)}{\Lambda_{W}r}\right\}+N\left\{\frac{\lambda-1}{\Lambda_{W}}f^{''}+\frac{2(1-\lambda)(f-1)}{\Lambda_{W}r^{2}}\right\}
\end{eqnarray}
They have two new solutions as
\begin{equation}
Q_{e}^{2}-Q_{m}^{2}=0,\; f=1-\Lambda_{W}r^{2}-\alpha
r^{\frac{2\lambda\pm\sqrt{6\lambda-2}}{\lambda-1}},\;N=ar^{-\frac{1+3\lambda\pm2\sqrt{6\lambda-2}}{\lambda-1}}\sqrt{f},
\end{equation}
where $\alpha,a$ are constants. The solutions has been got by paper
\cite{Lu2009}.

\section*{Acknowledgments}
The work was partially supported by NSFC Grant No.10535060,
10775002, 10975005 and RFDP. I would like to thank Bin Chen for
drawing my attention to \hl~ gravity and giving me many useful
suggestions.

\end{document}